\documentstyle[12pt]{article}

\begin{document}
\title{Symplectic connections, Noncommutative Yang Mills theory and Supermembranes}
\author{I. Martin $^1$ and A. Restuccia $^2$ \\ {\small Departamento de F\'{\i}sica, Universidad Sim\'on
Bol\'{\i}var, Venezuela} \\ {\small and} \\ {\small $^1$ Theoretical Physics 
Group, Imperial College , London University} \\ {\small $^2$ Department of Mathematics, 
King's College, London University.}}

\maketitle{\centerline{e-mail: isbeliam@usb.ve,isbeliam@ic.ac.uk, arestu@usb.ve}}
\begin{abstract}
    In built noncommutativity of supermembranes with central charges 
    in eleven dimensions is disclosed. This result is used to 
    construct an action for a noncommutative supermembrane where 
    interesting topological terms appear.  In order to do so, we first 
    set up a global formulation for noncommutative Yang Mills theory over general symplectic manifolds. We make the above constructions 
    following a pure geometrical procedure using the concept of 
    connections over Weyl algebra bundles on symplectic manifolds. The 
    relation between noncommutative and ordinary supermembrane actions is discussed.    

\end{abstract}

\section{Introduction}

The appearance of noncommutative geometry in the study of string 
theory was first seen in \cite{witnp1} where a noncommutative product was 
defined as overlap of half strings. This fact and subsequent work, 
including that of mathematicians using string techniques see 
\cite{doug} and references therein, led to the present 
believe that noncommutative geometry is necessarily relevant in the 
understanding of string physics. The best understood context in which 
noncommutative geometry arises is in the 
interaction of open strings with background fields, where  a  
constant background antisymmetric field is much larger than the background metric \cite{condou}. 

In \cite{witjh} the description of $Dp$-branes in terms of fields on a 
noncommutative space was analysed. In there  open strings were 
quantized in the presence of a constant $B$  background field. The precise limits, in which the 
noncommutativity appeared, were obtained when it was observed that the 
$\alpha'\leadsto 0$ limit allows to express the product of vertex 
operators as a noncommutative star product.

It is also important to understand from the point of view of the 
eleven dimensional $M$ theory how the noncommutativity arises. There has been 
several articles on the matter \cite{doug} mainly through M(atrix) 
theories, where the noncommutativity is related to  constant 
components of the three-antisymmetric tensor potential $C$. One would expect that the noncommutative description of $D$-branes 
could be obtained from a corresponding geometrical analysis of 
classical supermembranes without any need for a coupling to an external 
background field. After all, the supermembrane in the light cone gauge over a flat background 
shows as a gauge symmetry area preserving diffeomorphisms which 
correspond to symplectomorphisms in two dimensions. This result 
suggests that there should be a formulation of the supermembrane in 
terms of symplectic noncommutative gauge theories constructed from an intrinsic 
symplectic structure.  Also, such  noncommutative formulation of the supermembrane should correspond 
to the noncommutative construction drawn from the quantum analysis of the 
open strings by analogy with the case of the Born Infeld action. This action, describing $D$-branes for 
slow varying fields, may be obtained either from the one-loop analysis 
of strings or from a direct study of classical supermembranes  \cite{towns}. 

It is important to emphasize that the analysis on a constant $B$ 
field, which directly defines a symplectic two-form in the 
construction of the star product, may only be considered in a local 
approach since any symplectic two-form is locally diffeomorphic to a 
canonical form with constant coefficients (Darboux's Theorem). 
However,the corresponding global study  (with general 
symplectic two-forms not necessarily constant)is needed to provide insight 
into the structure of the action for noncommutative gauge field theories. 

In this article we will address the problem of noncommutativity arising 
intrinsically in the context of supermembranes in eleven dimensions as well as a more general 
framework for noncommutative Yang- Mills theories where global aspects 
are taken into account.

Essentially, to see any noncommutativity one has to look for a 
Poisson structure structure intrinsically defined in the supermembrane 
theory. In our work, for the sake of simplicity we will restrict the 
study to the case of symplectic manifolds which is enough for the 
discussion of the supermembrane and for most of the applications of noncommutative Yang-Mills,the more general formulation on 
Poisson manifolds \cite{konse} \cite{konscatt} will be treated elsewhere.

A non-degenerate symplectic two-form emerges naturally in 
supermembranes when non-trivial central charges are considered in the 
supersymmetric algebra. These charges imply the existence of a closed 
two-form with integral periods over the two dimensional spatial 
worldvolume. We may reinterpret this two-form as a Poisson 
structure over the worldvolume. In the case of the supermembrane with 
a non trivial central charge,  monopole type solutions were found for 
the minima energy levels of the Hamiltonian which provide proper non 
degenerate closed two-forms \cite{mrt}\cite{mor1}\cite{mor2},i.e.  symplectic structures over the spatial worldvolume. In this 
picture, the symplectic structure may be constant only on a Darboux 
chart. Hence, in order to have a correct and general formulation it 
is relevant to have a global construction of noncommutative gauge 
theories over symplectic manifolds with general symplectic two-forms.

One way to achieve this goal is to follow the original ideas of 
deformation quantization in \cite{mosq} and \cite{vey}. In the 
latter, the idea of glueing charts with Moyal type star products 
(constant $B$) was used. Later a more geometrical approach was 
introduced particularly in \cite{fedo}, see also \cite{asak}\cite{wess}. We will take results from \cite{fedo} to 
build in a geometrical formulation on a Weyl bundle, the 
noncommutative Yang-Mills connections and their corresponding action 
using a different point of view from \cite{asak}. 
Its projection to central terms of the bundle defined over the 
symplectic manifold (the world volume in the case of the 
supermembrane) provide the global construction of noncommutative Yang 
Mills theory. We will obtain explicit formulae where the presence of 
a symplectic connection appears manifestly together with the Yang- 
Mills potential. Having then this geometrical construction we may 
then apply it to the description of the $D=11$ supermembrane as a 
noncommutative gauge theory as shown in section 5. In section 2, we 
introduce needed geometrical concepts, definitions and notation in 
Weyl algebra bundles seeking to be selfcontained. In section 3, we construct the noncommutative 
Yang Mills theory for the symplectic `flat' case. In section 4 , 
previous results are extended to general symplectic manifolds, there 
new terms in the noncommutative Yang Mills action appear naturally 
associated to the symplectic curvature. In section 5, the 
noncommutative supermembrane action is constructed and its relation to 
the ordinary known supermembrane action with central charges is considered.

\section {Connections over the Weyl algebra bundle}

\noindent The purpose of this section is to introduce notation and definitions we 
will use in subsequent sections, for more details see \cite{fedo}.  We will consider 
here a symplectic manifold  ($\Sigma , \omega $) of dimension
$2n$. The two-form $ \omega $ defines a symplectic 
structure on each tangent space $T_{x}\Sigma$. The corresponding 
tangent bundle is $T\Sigma$. We will denote by $y^{\mu}¥$
the components of $y\in T_{x}\Sigma $, $\mu=1,....,2n$ , and $\omega_{\mu \nu}$ the
nondegenerate antisymmetric tensor defining the symplectic
structure over the fibers of $T\Sigma$. Also, we denote $x^{\mu}$ the local
coordinates over $\Sigma$, $\mu =1,...,2n$.

A formal Weyl algebra $W_x$ corresponding to the symplectic space 
$T_{x}\Sigma$ is an associative algebra  over the complex space $C$ 
with a unit, its elements being a formal series 

\begin{equation}
a(y,h)= \sum_{k, p \geq 0}¥h^k a_{k, \mu_{1}\ldots \mu_{p}} y^{\mu_{1}}\ldots  
y^{\mu_{p}}
\end{equation}
where h is a formal parameter, $\mu_{0} = 0$ and $\mu_{p}$ runs from $1$ to 
$2n$ when $p\neq 0$. To order terms in the summation, we give the 
following degrees to variables: deg $y^{\mu} = 1$, deg $h=2$ and we order 
by increasing degrees $2k + p$.
The Weyl product of elements $a,b\in W_x$ is defined as
\begin{equation}
a\circ b = \sum_{k=0}^{\infty}¥(- \frac{ih}{2})^{k} \frac{1}{k!}¥\omega^{{\mu_1}{\nu_1}}\ldots \omega^{{\mu_k}{\nu_k}}.
{\frac{\partial^{k} a}{\partial y^{\mu_{1}}\ldots \partial y^{\mu_{k}}}}
{\frac{\partial^{k} b}{\partial y^{\nu_{1}}\ldots\partial y^{\nu_{k}}}}
\end{equation}

This product is associative and independent of the basis in $T_{x}\Sigma $.The union of
$W_x$ defines the Weyl algebra bundle $W$. Sections of the Weyl bundle 
are functions
\begin{equation}
a(x,y,h)= \sum_{k, p \geq 0}¥h^k a_{k, \mu_{1}\ldots \mu_{p}}(x) y^{\mu_{1}}\ldots  
y^{\mu_{p}}
\end{equation}
where $a_{k, \mu_{1}\ldots \mu_{p}}(x)$ are symmetric covariant tensor 
fields on $\Sigma $.
We will also consider q-forms on $\Sigma$ with values in $W$,
\begin{equation}
a(x,y,h)= \sum h^k a_{k,\mu_{1}\ldots \mu_{p},\nu_{1}\ldots \nu_{q}}(x) 
y^{\mu_{1}}...y^{\mu_{p}}
dx^{\nu_{1}}\wedge\ldots\wedge  dx^{\nu_{q}}
\end{equation}
where coefficients are covariant tensor fields symmetric with respect
to $\mu_{1}\ldots \mu_{p}$ and antisymmetric in $\nu_{1}\ldots \nu_{q}$. The 
differential forms constitute an algebra with multiplication defined 
by means of the exterior product of differentials $dx^{\nu}$ and the 
Weyl product of polynomials in $y^{\mu}$. The commutator of two 
forms $a\in W\otimes\Lambda ^{q_{1}¥}¥$, $b\in W\otimes\Lambda 
^{q_{2}¥}¥$ is

\begin{equation}
[a,b]=a\circ b-(-1)^{q_1q_2}b\circ a.
\end{equation}
A central form $a$ is such that for any $b\in W\otimes\Lambda$, 
$[a,b]=0$

We now introduce  connections on the bundle $W\otimes\Lambda $.They 
are differential operators
\begin{equation}
{\cal {D}}: W\otimes\Lambda ^{q}\mapsto W\otimes\Lambda ^{q+1}
\end{equation}
such that for any scalar form $\phi\in\Lambda^{q}$ and section $a$ of $W\otimes\Lambda$ 
\begin{equation}
{\cal{D}}(\phi \wedge a)=d\phi\wedge a+(-1)^{q}\phi\wedge {\cal{D}}a 
\label{prop}
\end{equation}

To construct connections on the bundle $W\otimes\Lambda $, we will  
use the concept of  symplectic connections on $\Sigma $. There always 
exist a symplectic connection on any symplectic manifold. It is a 
torsion free connection that preserves the covariant tensor 
$\omega_{\mu\nu}$,i.e.
\begin{equation}
D_{\rho}\omega_{\mu\nu}=0,\label{sym}   
\end{equation}
$D_{\rho}$ being a covariant derivative with respect to the basis 
$\{\frac{\partial}{\partial x^{\rho}¥}\}$
 
The connection symbols  $\omega_{\mu\lambda}¥\Theta^{\lambda}_{\rho \nu} $ associated with the symplectic 
connection $D_{\rho}$ are completely determined by the defining equation (\ref {sym}) only up to an 
arbitrary completely symmetric tensor.     
\begin{equation}
\Theta_{\mu\rho\nu}=\frac{1}{3}\zeta_{\mu\rho\nu}+\frac{1}{3}\frac{\partial \omega_{\mu\nu}}{\partial x^{\rho}¥}+
\frac{1}{3}\frac{\partial \omega_{\mu\rho }¥}{\partial x^{\nu }¥}
\end{equation}
where $\zeta_{\mu\rho\nu}$ is completely symmetric and arbitrary.

We now lift the symplectic connections to the Weyl bundle, defining connections 
${\cal{ D}}_{S}$ on sections of $W\otimes\Lambda $ .They are
defined as
\begin{equation}
{\cal{ D}}_{S}a= dx^{\rho}\wedge D_{\rho}a   \label{der}
\end{equation}

The properties of the symplectic connection $D_{\rho}$ imply the 
following property of the connection ${\cal{ D}}_{S}$ on the Weyl 
product of the bundle $W\otimes\Lambda $
\begin{equation}
{\cal{ D}}_{S}(a\circ b)= {\cal{ D}}_{S}a\circ b + (-1)^{q_{1}} a\circ {\cal{ 
D}}_{S}b
\end{equation}
for $a\in W\otimes\Lambda ^{q_{1}¥}¥$ and $b\in W\otimes\Lambda 
^{q_{2}¥}¥$
Consequently they satisfy
\begin{equation}
{\cal {D}}_{S}[a,b]=[{\cal {D}}_{S}a,b]+(-1)^{q_{1}}[a,{\cal {D}}_{S}b].
\end{equation}

In Darboux local coordinates (coordinates where the components of the 
symplectic two-form $\omega$ are constants), the connection ${\cal{ D}}_{S}$ 
can be written as
\begin{equation}
{\cal {D}}_{S}a= da + \frac{i}{h}[\Theta ,a]    
\end{equation}
where $\Theta = \frac{1}{2} \Theta _{\mu\nu\rho}y^{\mu}y^{\nu}dx^{\rho}¥$

More general covariant derivatives ${\cal{ D}}$ on the bundle may be considered 
with one-form connections $\gamma$ globally defined  on $\Sigma $ and 
with values in $W$,
\begin{equation}
 {\cal {D}}a={\cal {D}}_{S}a + \frac{i}{h}[\gamma ,a]   
\end{equation}
The two-form $ \Omega$
\begin{equation}
    \Omega =R + {\cal {D}}_{S}\gamma + \frac{i}{2h}[\gamma,\gamma]
\end{equation}
is the Weyl curvature of the connection ${\cal {D}}$.  $R$ is the 
curvature of the  connection ${\cal {D}}_{S}$. The Weyl 
curvature satisfies the Bianchi identity
\begin{equation}
    {\cal {D}} \Omega ={\cal {D}}_{S}\Omega  + 
    \frac{i}{h}[\gamma,\Omega]=0 
\end{equation}
moreover, for any section $a\in W\otimes\Lambda ¥$
\begin{equation}
   {\cal {D}}^{2}a = \frac{i}{h}[\Omega,a]  
\end{equation}

In general, transitions on the bundle $T\Sigma$ will induce transitions 
on the algebra $W$.The infinitesimal
gauge transformations on elements of the algebra are expressed as 
automorphisms given by
\begin{equation}
a\rightarrow a+[a,\lambda]
\end{equation}
with `infinitesimal' $ \lambda\in W$.\\ The corresponding gauge
transformations for the  connections  ${\cal {D}}$ are
\begin{equation}
{\cal {D}}\rightarrow{\cal {D}}+{\cal {D}}\lambda.
\end{equation}
Consequently
\begin{equation}
{\cal {D}} a\rightarrow{\cal {D}} a +[{\cal {D}} a,\lambda].
\end{equation}

Abelian connections ${\cal {D}}_{A}$ are connections ${\cal {D}}$ with  Weyl curvature $\Omega$
being a central form of the algebra. Let us denote it $\Omega_A$. It
then satisfies
\begin{equation}
[\Omega_A,a]=0
\end{equation}
for any section $a \in W$.

There always exist abelian connections on $W$, an example of these 
connections may be expressed as
\begin{equation}
{\cal {D}}_{A}={\cal 
{D}}_{S}+\frac{i}{h}[\omega_{\mu\nu}y^{\mu}dx^{\nu}+r, \cdot] 
\label{echa}
\end{equation}
with deg $r$ $\ge3$ \cite{fedo}.
Associated with ${\cal {D}}_{A}$ there is a subalgebra of
$W$, denoted $W_{A}$, defined by
\begin{equation}
W_{A}¥=\Big\{a\in W: {\cal {D}}_{A}a=0\Big\}.
\end{equation}
There is a one to one correspondence between the $C^\infty$
functions $a_{0}(x)$ over $ \Sigma$ and the elements of $W_{A}$. In
fact, given $a\in W_{A}$ , one defines the projection
\begin{equation}
\sigma a:=a(x,y=0,h)=a_0(x),
\end{equation}
and given $a_{0}(x)$ there is a unique element $a\in W_{A}$ with
such projection. The explicit expression of this element, obtained
by solving ${\cal {D}}_{A}a=0$, is
\begin{eqnarray}
a(x,y,h)=&& a_0(x)+D_{\mu}a_{0}(x).y^{\mu}+
\frac{1}{2} D_{\mu}D_{\nu}a_{0}(x) y^{\mu}y^{\nu}+
\frac{1}{6}D_{\mu}D_{\nu}D_{\rho} 
a_{0}(x)y^{\mu}y^{\nu}y^{\rho}\nonumber \\ && -
\frac{1}{24} R_{\mu\nu\rho\lambda}\omega^{\lambda \sigma}D_{\sigma}a_{0}y^{\mu}y^{\nu}y^{\rho}+\ldots
\end{eqnarray}
where the remaining terms are of higher degree.
If $a$ and $b$ $\in W_{A}$, then
\begin{equation}
\sigma (a\circ b)=a_{0}\star b_{0}
\end{equation}
where $\star$ is the globally defined star product. In the
particular case when $\omega_{\mu\nu}$ is constant and the symplectic 
connection is zero, the formula agrees with the Moyal
product.

Finally, we may obtain the above constructions also for  more 
general  Weyl bundles where the fibre $T_{x} \Sigma$ is replaced by 
$L_{x}$. The latter being any symplectic vector space on $\Sigma$ of 
dimension $2n$ with a given symplectic structure $\omega$ and 
corresponding symplectic connection $D_{L}$. The bundle $L$ is assumed 
to be isomorphic to $T \Sigma$ where the bundle isomorphism is
\begin{equation}
    \varepsilon : TM\rightarrow L
\end{equation}
Introducing a local symplectic frame $(e_{1}\ldots e_{2n})$ of $L$ and a 
dual frame $(e^{1}\ldots e^{2n})$ for the dual $L^{*}$, we will have 
local one-forms on $\Sigma$ $ \varepsilon^{*}(e^{i})= 
\varepsilon^{i}_{\mu} dx^{\mu}¥$ corresponding to a basis $\{dx^{\rho}\}$ of $T^{*}\Sigma$. The form $\omega$ 
in $L$ may be transported to $T\Sigma $ giving a nondegenerate 
two-form on $\Sigma$
\begin{equation}
    \omega = \frac{1}{2}\omega_{ij} e^{i} \wedge e^{j} = 
    \frac{1}{2}\tilde{\omega}¥_{\mu\nu} dx^{\mu} \wedge dx^{\nu}
\end{equation}
Once the isomorphism between $L$ and $T\Sigma $ is defined the corresponding 
concepts like symplectic and abelian connections apply as above.

\section{Yang-Mills connections over the Weyl bundle}

We will consider in this section one-form connections and curvatures 
on $u(1)$ valued Weyl bundles over a manifold $\Sigma$ with a symplectic two-form 
whose coefficients are constants, in this 
case we take $\varepsilon^{i}_{\mu}=\delta ^{i}_{\mu}$. We will 
assume also that the totally symmetric symplectic one-form 
connection $\Theta$ is zero. These assumptions will be relaxed in the 
next section. They are valid, for example, on the symplectic vector 
space $R^{2n}$ with a constant symplectic two-form $\omega$. The extension 
to the $u(N)$ non abelian case may be achieved by 
considering the Weyl algebra bundle with elements $a$ valued in the 
$u(N)$ algebra. The following construction may then be extended in a 
straightforward way.

We denote by $\Omega_{A}^{0} (\Sigma ,W)$ the set of sections $a(x,y)$ of $W$ 
satisfying
\begin{equation}
    {\cal {D}}_{A}a(x,y)=0
\end{equation}
where ${\cal {D}}_{A}$ in (\ref{echa}) with $r=0$ takes the particular form 
\begin{equation}
    {\cal {D}}_{A}a(x,y)= da + \frac{i}{h}[\epsilon_{ij}y^{i}dx^{j}, 
    a],
\end{equation}
and $\Omega_{A}¥^{1} (\Sigma ,W \otimes \Lambda ^{1})$ the set of sections 
$b= e^{i} b_{i}(x,y)$ of Weyl algebra valued one-forms in $W \otimes \Lambda 
^{1}$ in a particular basis  $(e^{i}\ldots e^{2n})$ that fulfill the 
condition
\begin{equation}
    {\cal {D}}_{A}b_{i}(x,y)=0
\end{equation}
the coefficients $b_{i} \in \Omega_{A}^{0} (\Sigma ,W) $. Any 
elementc $a \in \Omega_{A}^{0} (\Sigma ,W)$ satisfy the equation
\begin{equation}
    a=  a_{0} + \delta^{-1}({\cal {D}}_{A} + \delta )a(x,y) \label{iter}
\end{equation}
where ${\cal {D}}_{A}$ is the abelian connection of section 2 and
\begin{equation}
 \delta = e^{i}\frac{\partial}{\partial y^{i}},      
\end{equation}
$\delta^{-1}$ is an operator acting on every term $c_{pq}= 
c_{i_{1}\ldots i_{p},\j_{1}\ldots j_{q}}y^{i_{1}}\cdots 
y^{i_{p}}e^{j_{1}}\cdots e^{j_{q}}$ of $c(x,y) \in W\otimes\Lambda$ as follows
\begin{equation}
 \delta^{-1}c_{pq}= \frac{1}{p+q}y^{k} \imath (e_{k})c_{pq}   
\end{equation}
here $\imath (e_{k})$ is the contraction on one-forms operation. These 
operators satisfy a formula similar to the Hodge decomposition
\begin{equation}
    c(x,y) =  \delta \delta^{-1}c + \delta^{-1}\delta c + c_{00}(x)
\end{equation}
where $c_{00}(x)$ is the zero-form central term of $c$. Notice that $\delta$ 
decreases the degree of every term of $c$ in one, while $\delta^{-1}$ 
increases it in one.

Given the first term $a_{0}(x)$ in (\ref {iter}), the rest of the terms in the series 
$a(x,y)$ can be calculated using an iterative procedure by applying 
successively the same equation (\ref {iter}).

We introduce now the lifted symplectic connections and curvatures on $W$ 
whose projections to central terms yield the usual noncommutative 
Yang-Mills field strengths.

We consider first the connection $\hat{{\cal {D}}}¥$ given explicitly by

\begin{equation}
    \hat{{\cal {D}}}= d + \frac{i}{h}[\gamma,\bullet]
\end{equation}
locally  $\gamma$ is in $\Omega_{A}¥^{1} (\Sigma ,W \otimes \Lambda ^{1})$.

If $a \in \Omega_{A} ^{0} (\Sigma ,W )$, then
\begin{equation}
    \hat{{\cal {D}}}a= e^{i}(\partial_{i}a + \frac{i}{h}[\gamma_{i},a]) \in 
    \Omega_{A}¥^{1},
\end{equation}
its projection yields
\begin{equation}
    \sigma \hat{{\cal {D}}}a= e^{i}(\partial_{i}a_{0}(x) + \frac{i}{h} \{{\cal 
    {A}}_{i}¥,a_{0}\}_{Moyal})  \label{proj}
\end{equation}
where $\sigma\gamma_{i}={\cal {A}}_{i}$.

Condition $\gamma \in \Omega_{A}¥^{1} (\Sigma ,W \otimes \Lambda ^{1})$ 
is necessary to get the desired projection over $\Lambda ^{1}(\Sigma 
)$ in ( \ref {proj}). The more general condition ${\cal {D}}\gamma= 0$ 
is not enough to achieve the required projection.

We define the infinitesimal gauge variations $ \Delta_{g} $ of sections $a \in 
\Omega _{A}^{0} (\Sigma ,W)$ as
\begin{equation}
   \Delta_{g}   a(x,y)= [a,\lambda ] 
\end{equation}
where the infinitesimal parameter $\lambda\in \Omega _{A}^{0} (\Sigma ,W)$.
We then define
\begin{equation}
     \Delta_{g}   \gamma(x,y)= \hat{{\cal {D}}} \lambda , \label{gamm}
\end{equation}
implying
\begin{equation}
    \Delta_{g}  \hat{{\cal {D}}}a= [\hat{{\cal {D}}}a, \lambda].
\end{equation}
In particular, we obtain from equation (\ref {gamm})
\begin{equation}
     \Delta_{g} {\cal {A}}_{i}= \partial_{i}\lambda_{0} + \frac{i}{h}\{{\cal 
    {A}}_{i}(x),\lambda_{0}(x)\}_{Moyal} 
\end{equation}
where $\sigma\lambda(x,y) =\lambda_{0}(x)$.

We remark that $\Delta_{g} \gamma \in \Omega_{A}¥^{1} (\Sigma ,W 
\otimes \Lambda ^{1})$, this is a non-trivial property of the construction. In fact, 
if the curvature of the symplectic connection was not zero , 
then $e^{i}\partial_{i}\lambda$ would not belong to $\Omega_{A}¥^{1}$ and the above 
construction had to be modified so that all gauge equivalence classes 
could belong to $\Omega_{A}¥^{1}$. We notice that ${\cal {D}}_{A}$ 
corresponds to the zero in the space of connections we are 
considering and it is invariant under infinitesimal gauge 
transformations. This result implies that $\hat{{\cal {D}}} - {\cal 
{D}}_{A}$ defines the same connection as $\hat{{\cal {D}}}$ on any $\Omega_{A}¥^{q} (\Sigma ,W \otimes 
\Lambda ^{q})$. So in considering physical connections we will take 
those ones normalized to zero, in this case ${\cal {D}}= \hat{{\cal {D}}} - {\cal {D}}_{A}$ besides, this 
connection representation is more appropriate to study global aspects 
of the bundle.

We may now write the curvature of the connection ${\cal {D}}$,
\begin{equation}
  \Omega= d\gamma + \frac{i}{2h}[\gamma,\gamma] - \omega,
\end{equation}
it satisfies the Bianchi identity
\begin{equation}
    {\cal {D}}\Omega= 0,
\end{equation}
under infinitesimal gauge transformations, we obtain
\begin{equation}
    \Delta_{g} \Omega = [\Omega, \lambda].
\end{equation}

The projection of $\Omega$ becomes the usual noncommutative Yang-Mills 
field strength
\begin{equation}
    \sigma \Omega=\frac{1}{2} e^{i}\wedge  e^{j}(\partial_{i}{\cal 
    {A}}_{j} - \partial_{j}{\cal {A}}_{i}  + \frac{i}{h}\{{\cal 
    {A}}_{i},{\cal{A}}_{j}\}_{Moyal}) - \omega \label{cur} = {\cal 
    {F}} - \omega
\end{equation}
equation (\ref {cur}) corresponds to $u(1)$ noncommutative Yang-Mills 
theory.  We may now define the $u(1)$ Yang-Mills action 
over the Weyl algebra bundle
\begin{equation}
    S_{YM} = \int_{\Sigma} \sigma (\Omega \circ *\Omega) \label{ymom}
\end{equation}
where $*\Omega \in W\otimes\Lambda ^{2n-2}$ is the Hodge dual to $\Omega$ 
constructed using the induced metric $ g(\cdot,\cdot )$determined by a compatible complex structure $J$ 
and the symplectic structure $\omega (\cdot,\cdot )$ on the vector bundle $L$,
\begin{equation}
    g(u, v) = \omega (u, Jv) \hspace{3pt}{\mathrm{for} \hspace{3pt}\mathrm{any} \hspace{3pt}\mathrm{two}\hspace{3pt}
    \mathrm{vectors}} \hspace{3pt} u,v \in  L_{x},
\end{equation}
complex structures $J$ always exist for any symplectic vector bundle. 
Equation (\ref{ymom}) may be written as  
\begin{equation}
      S_{YM}= \int_{\Sigma} {\cal {F}}\wedge *{\cal {F}} 
      -2\int_{\Sigma}{\cal {F}}\wedge *{\omega} + Vol_{{\Sigma}}
\end{equation}

\section{Global construction of noncommutative gauge theories on the 
Weyl algebra bundle }
Let $\Sigma $ be a symplectic manifold with a symplectic two-form 
$\omega_{\mu\nu}¥ dx^{\mu}\wedge dx^{\nu}$. In 
this section, we assume $\omega$ to be an arbitrary non-degenerate closed 
two-form over $\Sigma $.
A set of multi-beins is defined by
\begin{equation}
\omega_{\mu\nu} = 
\varepsilon_{\mu}¥^{i}\varepsilon_{\nu}¥^{j}\epsilon_{ij}, \label{ome}    
\end{equation}
where $\epsilon_{ij}$ is the canonical symplectic tensor. Because of Darboux 
theorem, locally we always have
\begin{equation}
    \varepsilon_{\mu}¥^{i}= \partial_{\mu} g^{i}.  \label{vec}
\end{equation}¥

We may consider an atlas where on each chart we have (\ref{vec}) . The 
transitions on $g^{i}$ between different charts preserve the symplectic 
structure (\ref{ome}). The multi-bein $\varepsilon_{\mu}¥^{i}$ will then 
have transitions over $\Sigma$, otherwise, one would have a set of $2n$ non-singular vector fields 
globally defined over $\Sigma$, but this is not true in general.

Let us discuss the transitions on intersection of charts in more 
detail. Consider two open sets $U$ and $\hat{U}$, $U\bigcap \hat{U} =\emptyset$ in which
\begin{equation}
 \varepsilon_{\mu}¥^{i}= \partial_{\mu} g^{i}, \hspace{3pt} 
\mathrm{and} \hspace{3pt}¥\hat{\varepsilon}_{\mu}¥^{i}= \partial_{\mu} \hat{g}^{i}       
\end{equation}
respectively. In $U\cap  \hat{U}$ we then have
\begin{equation}
 \omega_{\mu\nu} = 
\varepsilon_{\mu}¥^{i}\varepsilon_{\nu}¥^{j}\epsilon_{ij}¥= 
\hat{\varepsilon}_{\mu}¥^{i}\hat{\varepsilon}_{\nu}¥^{j}\epsilon_{ij},   
\end{equation}
from which we obtain
\begin{equation}
\hat{\varepsilon}_{\mu}¥^{i}= S_{j}^{i}\varepsilon_{\mu}¥^{j} \hspace{6pt} 
{\mathrm{where}}\hspace{3pt} 
S_{j}^{i} = 
\epsilon^{ik}\hat{\varepsilon}_{k}¥^{\mu}\varepsilon_{\mu}¥^{l}\epsilon_{lj}, \label{veco}     
\end{equation}¥
we define the inverse of $\epsilon$ by $\epsilon^{ij}\epsilon_{jk}=\delta_{k}^{i}$.
One may verify that $S$ preserves the canonical symplectic tensor and 
hence $S\in Sp(2n)$.
Consequently in order to have a global construction over $\Sigma$, one must 
begin  by introducing a symplectic $Sp(2n)$ connection on the tangent bundle.
We first consider the following symplectic connection over $\Sigma$,
\begin{equation}
    \Theta_{\mu\nu}^{\lambda}\equiv 
    \zeta_{\mu\nu\rho}\omega^{\rho\lambda} 
    +\frac{1}{3}\frac{\partial\omega_{\nu\rho}¥}{\partial 
    x^{\mu}¥}\omega^{\rho\lambda} + 
    \frac{1}{3}\frac{\partial\omega_{\mu\rho}}{\partial x^{\nu}}\omega^{\rho\lambda} \label{cosy3}
\end{equation}¥
where $\zeta_{\mu\nu\rho}$ is a totally symmetric tensor. This is the most general 
expression for a connection satisfying
\begin{equation}
    (\frac{\partial}{\partial x^{\mu}} + \Theta_{\mu})\omega_{\rho 
    \lambda }=0
\end{equation}
here $ \Theta_{\mu\nu}^{\lambda}$ is expressed in terms of $\omega_{\mu\nu}$ and its derivatives. It is 
invariant under the $Sp(2n)$ transition of the multi-bein.
We now consider the following torsion free connection on the tangent 
space¥
\begin{equation}
     \Gamma_{\mu i}^{j}= 
    \varepsilon_{i}¥^{\nu}(\frac{\partial\varepsilon_{\nu}¥^{j}}{\partial 
    x^{\mu}¥} -  \Theta_{\mu\nu}^{\lambda}\varepsilon_{\lambda}¥^{j})  \label{cosy2}  
\end{equation}
it transforms as a $Sp (2n)$ connection under  $Sp(2n)$ transformations on the 
tangent space. In fact,¥
\begin{equation}
    \hat{\Gamma}_{\mu i}^{j} = (S^{-1})_{i}^{l} \Gamma_{\mu 
    l}^{k}S_{k}^{j} - 
    (S^{-1})_{i}^{k}\frac{\partial S_{k}^{j}}{\partial x^{\mu}}
\end{equation}¥
This connection is symplectic on the tangent space.  We may construct 
from it the most general symplectic connection on the tangent space in the 
following way. Let us denote 
\begin{equation}
  \breve{D}_{\mu} \equiv \partial_{\mu} + \Gamma_{\mu}, \label{cosy}¥
\end{equation}
a symplectic connection must satisfy
\begin{equation}
 (\breve{D}_{\mu} + \Delta\Gamma_{\mu})\epsilon_{ij} = 0     
\end{equation}¥
this equation has the general solution
\begin{equation}
 \Delta \Gamma_{\mu i}^{j}= 
    \frac{1}{3} (\breve{D}_{\mu}\epsilon_{il})\epsilon^{lj} +    
  \frac{1}{3}\varepsilon_{\mu}¥^{k}\varepsilon_{i}¥^{\nu}(\breve{D}_{\nu}\epsilon_{kl})\epsilon^{lj} +  
  \varepsilon_{\mu}¥^{k}\tilde \zeta_{(ilk)} \epsilon^{lj} 
  \label{cosy1}   
\end{equation}¥
$\Delta \Gamma_{\mu i}^{j}$ is a covariant vector on the world volume and a tensor 
under $Sp(2n)$ transformations. Since the connection (\ref{cosy}) is symplectic 
the first two terms of the right hand side member in (\ref{cosy1}) are zero. We may finally construct our symplectic connection $D$ , when acting on mixed indices vectors $V^{i}_{\nu}$ it yields
\begin{equation}
    D_{\mu}V_{\nu}^{i} = \frac{\partial V_{\nu}^{i}}{\partial x^{\mu}} + (\Gamma_{\mu} + 
     \Delta\Gamma_{\mu })_{l}^{i} V_{\nu}^{l} 
     -\Theta_{\mu\nu}^{\lambda}V_{\lambda}^{i},\label{str}
\end{equation}¥
it satisfies
\begin{equation}
     D_{\mu}\omega_{\rho\lambda}=0 ,\hspace{9pt}  D_{\mu}\epsilon_{ij}=0
\end{equation}
and it has the right transformation law on the world volume and in 
the tangent space. If we impose $\Delta \Gamma_{\mu i}^{j}$ to be 
zero, that is if we take the totally symmetric term in (\ref{cosy1}) zero, then 
the symplectic connection (\ref{str}) acting on the multibein is zero. This 
property is valid for any totally symmetric symbol in (\ref{cosy3}). The 
$Sp(2n)$ symplectic connection reduces in this case to (\ref{cosy2}). In the evaluation of the final formulas we will consider this kind of 
connection. We will call $D$ an $Sp(2n)$ symplectic connection,  when it acts on geometrical objects on the tangent space only.

We will now introduce a connection ${\cal {D}}$ on $W$ which in the flat 
limit when  $\omega$ has constant coefficients, $\varepsilon_{\mu}¥^{k}=\delta_{\mu}¥^{k}$ 
and $\Theta_{\mu\nu}^{\lambda}= 0$ reduces to the Yang-Mills connections of section 3. It will be a map 
from $\Omega_{A}¥^{0} \mapsto \Omega_{A}¥^{1}$, which in addition to the 
general property of any connection (\ref{prop}) it satisfies the Leibnitz property for the Weyl product in $W$. All 
these properties should, of course, be preserved under gauge transformations. We consider
\begin{equation}
    {\cal {D}} = \frac{i}{h}[G_{i}e^{i},\bullet] + \frac{i}{h}[\gamma,\bullet] 
    \label{conn}
\end{equation}¥
where $G_{i}$ obeys the following equations
\begin{equation}
  {\cal {D}}_{A}G_{i} = 0  , \hspace{3pt}  \sigma G_{i}= \epsilon_{ij} g^{j}(x) 
   \label{ge}
\end{equation}¥
where $g^{j}(x)$ is defined in (\ref {vec}) and $\gamma_{i}$ also obeys  
\begin{equation}
 {\cal {D}}_{A}\gamma_{i}= 0 .     \label{symm}
\end{equation}¥
recalling equations (\ref{der}) and (\ref{echa}), ${\cal {D}}_{A}$ is now written using the symplectic 
connection in (\ref{str})
\begin{equation}
{\cal {D}}_{A}={\cal {D}}_{S}+\frac{i}{h}[\epsilon _{ij}y^{i}e^{j}+r, \cdot]
 \end{equation}
We will denote the projection of $\gamma_{i}$ as ${\cal {A}}_{i}(x)$, like in section 3
\begin{equation}
   \sigma\gamma_{i}= {\cal {A}}_{i}(x) \label{cono} 
\end{equation}

\noindent ${\cal {D}}$ has the folowing properties:
\vspace{4pt}

\noindent a) If $a\in\Omega_{A}^{0}$ then ${\cal {D}}a\in\Omega_{A}^{1}$. This is so because the abelian connection ${\cal {D}}_{A}$ acts 
directly inside the bracket, and then equations (\ref{ge}) and 
(\ref{symm}) assure that it annihilates all terms inside it.
\vspace{3pt}

\noindent b) If $a\in\Omega_{A}^{q}$ then
\begin{equation}
 {\cal {D}}(a\circ b)=  {\cal {D}}a\circ b + (-1)^{q} a\circ {\cal {D}}b   
\end{equation}
because the bracket has that property.
\vspace{3pt}

\noindent c) If $a\in\Omega_{A}^{0}$,
\begin{equation}
    \frac{i}{h} \sigma [G_{i}e^{i},a]= 
    e^{i}\epsilon^{kl}\partial_{k} g^{j}\epsilon_{ij}\partial_{l}a_{0}  + O(h) = 
    e^{i}\partial_{i}a_{0} + O(h).
\end{equation}¥
since
\begin{equation}
\partial_{k}g^{i} = \varepsilon_{k}^{\mu}\partial_{\mu} g^{j}= 
\delta_{k}^{i}    
\end{equation}
the terms $O(h)$ depend on the curvature of the symplectic connection 
(\ref{str}) and becomes zero in the flat limit.
Consequently, in that limit $[G_{i}e^{i},a]$ is the element of $\Omega_{A}^{1}$ with 
projection $ e^{i}\partial_{i}a_{0}$. It then coincides with ${\cal{ 
D}}_{S}a$, which in the flat limit (and 
only there) is also an element of $\Omega_{A}^{1}$. We conclude that in the flat 
limit (\ref{conn}) is exactly the Yang-Mills connection of section 3.
\vspace{3pt}

\noindent d) We define the gauge transformations in a chart by
\begin{equation}
    \Delta_{g}\gamma = {\cal {D}}\lambda
\end{equation}¥
where $\lambda \in \Omega _{A}^{0}¥$ is the infinitesimal gauge parameter 
as before. It then preserves the 
form of (\ref {conn}).
\vspace{3pt}

\noindent e) The curvature of the connection ${\cal {D}}$ is then given by
\begin{equation}
    \Omega = \frac{i}{2h} [G,G] + \frac{i}{h} [G,\gamma] +  
    \frac{i}{2h} [\gamma,\gamma],\label{curva}
    \end{equation}¥
it satisfies the Bianchi identity
\begin{equation}
  {\cal {D}} \Omega =0   
\end{equation}¥
this property follows from the Jacobi identity for the bracket.
The first term in (\ref{curva}) reduces in the flat limit to
\begin{equation}
    \frac{i}{2h} [G,G] = - \frac{1}{2}¥e^{i}\wedge e^{j} \epsilon_{ij} = -  
    \omega  
\end{equation}¥
The projection of $ \Omega$ has in general the expression
\begin{eqnarray}
    \sigma \Omega =&& -\omega  + {\cal {F}} - 
   \frac{h^{2}}{96} \left(R_{jkli}(D_{\hat{j}}D_{\hat{k}}D_{\hat{l}} {\cal {A}}_{m} 
    -\frac{1}{4}R_{{\hat{j}}{\hat{k}}{\hat{l}}p}\epsilon^{pq}D_{q}{\cal {A}}_{m})\right)\epsilon^{j\hat{j}}\epsilon^{k\hat{k}}\epsilon^{l\hat{l}}e^{i}\wedge e^{m} \nonumber \\ && - \frac{h^{2}}{96\cdot 8}R_{jkli}R_{{\hat{j}}{\hat{k}}{\hat{l}}m}\epsilon^{j\hat{j}}\epsilon^{k\hat{k}}\epsilon^{l\hat{l}}e^{i}\wedge e^{m}  + O(h^{3})              \ldots
\end{eqnarray}¥
where the curvature is constructed from the $Sp(2n)$ symplectic connection 
(\ref{str}), the remaining terms are higher order in $h$ and depend 
also on the derivatives of the curvature. 
The curvature ${\cal {F}}$ is the Yang Mills field strength 
\begin{equation}
    {\cal {F}}= \frac{1}{2}e^{i}\wedge e^{j}( D_{i}{\cal 
    {A}}_{j}-D_{j}{\cal {A}}_{i} + \frac{i}{h}\{{\cal 
    {A}}_{i},{\cal {A}}_{j}\}_{star})\label{efe}
\end{equation}¥
constructed now with the $Sp(2n)$ covariant symplectic derivative 
introduced in (\ref{str}) , notice that 
the $star$ bracket in (\ref{efe}) is the global generalization of the Moyal 
bracket over the whole symplectic manifold obtained in \cite{fedo}, 
briefly presented in section 2. We notice that, because of (\ref{vec}) 
and (\ref{veco}), the first  covariant symplectic derivative of $g^i$
is a simple derivative. We will  assume the same transformation law under 
$Sp(2n)$ for ${\cal {A}}$ in  (\ref{cono}). As in the previous section the above 
construction may be extended to $u(N)$ valued  Weyl algebras in a straightforward way.

\section{Supermembranes and noncommutative gauge theories}

As already said before, the supermembrane in the light cone gauge 
shows as a gauge symmetry  symplectomorphisms in two dimensions. This 
result  lead us to look for a formulation of the supermembrane in 
terms of symplectic noncommutative gauge theories constructed from an intrinsic 
symplectic structure.

The starting point in the construction we have discussed in previous 
sections is the symplectic manifold $\Sigma$. From the global symplectic 
two-form $\omega$ we constructed all the geometric objects which allow to 
formulate Yang-Mills connections over the Weyl algebra bundle. After 
projection we obtained a global noncommutative formulation of 
Yang-Mills connections. For the case of the supermembrane in the 
light cone gauge, the world volume $\Sigma$ is a Riemann 
surface. But now, how do we incorporate a symplectic structure into it?. A 
natural way is to consider a supermembrane with a nontrivial central charge 
of the SUSY algebra \cite{town2}. That is, consider $Z^{12}$ 
\begin{equation}
Z^{12}= \int_{\Sigma} dX^{1}\wedge dX^{2} = 2\pi n \label{weil}   
\end{equation}
which requires that when ${\Sigma}$ is a compact Riemann surface,  
$X^{1}$ and $X^{2}$ must be winded on compactified directions. By  the 
Weil theorem,there always exists a nondegenerate closed two-form  $\omega$ 
satisfying a condition like (\ref {weil}). In particular, we may 
consider a solution for the winded case on a compact Riemann surface $\Sigma$ where 
the Hodge dual of $\omega$ is an integer, 
\begin{equation}
    *\omega=n
\end{equation}
over all $\Sigma$, the monopole solution. The density used in $*\omega$ is the one 
introduced in the light cone gauge fixing procedure. This monopole 
configuration has a natural generalization for other p-branes in 
terms of extended self-dual connections \cite{mor1} \cite{mrt} \cite{br2}.

We may thus introduce in an intrisic way a nondegenerate closed 
two-form $\omega =  dX^{1}\wedge dX^{2}$ over $\Sigma$. This closed two-form is invariant under the area 
preserving diffeomorphism \cite{mrt} which is the residual gauge 
symmetry on the supermembrane in the light cone gauge. We may now construct, using 
our approach of the previous sections, noncommutative Yang-Mills 
connections globally defined on the Weyl bundle over $\Sigma$.

We consider the seven transverse coordinates to the supermembrane as 
projections of corresponding elements $X^{M}(x,y), M= 1,\ldots 7$  belonging to $\Omega_{A}^{0}$. The 
gauge transformations of $\sigma X^{M} $ with respect to the area preserving 
diffeomorphisms on the membrane correspond exactly to the gauge 
transformations constructed over the Weyl algebra bundle for the 
elements of $\Omega_{A}^{0}$. These elements will have associated 
conjugate momenta densities $P^{M}$, we will also denote $\acute{P}^{M}$ 
the associated scalar fields $\acute{P}^{M} = 
\frac{1}{\sqrt{det\omega}}P^{M}$ and $\acute{P}^{M}\in \Omega_{A}^{0}$. Also, we consider 
$\gamma=\gamma_{i}e^{i}, i=1,2$ and $\Pi = {\Pi}^{i}\omega_{ij}e^{j} $
the one-form gauge field and its corresponding momentum density, respectively. 
We denote $\acute{\Pi}= \frac{1}{\sqrt{det\omega}} 
{\Pi}$ the associated one form momemtum. Notice that 
$\acute{\Pi}$  belong to 
$\Omega_{A}^{1}$. The symplectomorphisms are generated by the first class constraint
\begin{equation}
{\cal {D}}_i{\Pi}^{i}+ [X^{M},P_{M}] = 0    
\end {equation}
All elements $X^{M}$, $\acute{P}^{M}$ and ${\acute{\Pi}^{i}}$ will 
transform homogeneously under an  infinitesimal gauge transformation 
with parameter $\lambda$ as in the previous section
\begin{equation}
\Delta_{g} \cdot = [\cdot, \lambda]    
\end {equation}
also, the gauge field will transform accordingly as
\begin{equation}
\Delta_{g}\gamma = {\cal {D}}\lambda
\end {equation}
where ${\cal {D}}$ is the connection introduced in section 4.

We may then write the following Hamiltonian density on the Weyl 
algebra over $\Sigma$, subject to the above first class constraint, as a noncommutative Yang Mills coupled to 
the scalar fields representing the tranverse coordinates to the membrane,
\begin{eqnarray}
{\cal H} =&& \frac{1}{2}(\acute{P}^{M} \circ *\acute{P}^{M})+  \frac{1}{2}(\acute{\Pi} \circ 
*\acute{\Pi}) -  \frac{1}{2 h^{2}}([G_{i}e^{i} + \gamma ,X^{M}] \circ 
*[G_{i}e^{i} + \gamma ,X^{M}])   \nonumber \\  && - 
\frac{1}{4 h^{2}}([X^{M},X^{N}] \circ *[X^{M},X^{N}]) +  
 \frac{1}{2} (\Omega\circ *\Omega) 
\end{eqnarray}

We may inmediately project out the center components of this 
Hamiltonian yielding
\begin{eqnarray}
\sigma {\cal H} =&& \frac{1}{2}(\acute{P}^{M})^{2}\omega + \frac{1}{2}(\acute{\Pi}\wedge 
*\acute{\Pi}) -  (D_{{\cal{A}}} X^{M} \wedge *D_{{\cal{A}}} X^{M})  \nonumber 
\\  && - 
  \frac{1}{4 h^{2} } \{X^{M},X^{N}\}_{star}^{2}\omega + \frac{1}{2}({\cal{F}} -  
   \omega)(*{\cal{F}} -  *\omega )   +  
   \mathrm{curvature}\hspace{3pt} \mathrm{terms}  
\end{eqnarray}
where $D_{{\cal{A}}}=  D_{S} X^{M} + \{{\cal{A}}, X^{M} \}_{star}$ corresponds to the first terms in a projection of the gauge 
covariant derivative,
\begin{eqnarray} 
\sigma{\cal{D}}_{i}X^{M} =&&  
\frac{i}{h} \{G_{i},X^{M} \}_{star} +  \frac{i}{h} \{{\cal{A}}_{i}, X^{M} \}_{star} \nonumber 
 \\  =&& D_{S\hspace{2pt}i} X^{M} + 
 \frac{i}{h} \{{\cal{A}}_{i}, X^{M} \}_{star} +  \mathrm{curvature}\hspace{3pt} \mathrm{terms}  \ldots
\end{eqnarray}

All constructions of sections 2, 3 and 4 are based on properties 
related to the Weyl bracket, they do not rely on the Weyl product by itself. 
These two basic properties are the Jacobi identity and that  
covariant derivatives satisfy the Leibniz condition with respect to 
the bracket. We may reconstruct everything if we replace the Weyl 
bracket by a Poisson bracket on the Weyl bundle. That is, we may 
change the noncommutative gauge symmetry based on a Weyl bracket in the theory for a 
symplectic gauge symmetry based on a Poisson bracket.  The symplectic 
connections fulfill again the Leibniz condition with respect to it. We 
may then proceed to define the connections
\begin{equation}
    {\cal{D}} = D_{S} + [\gamma,\bullet]_{P}
\end{equation}
where now
\begin{equation}
    [\cdot,\cdot]_{P} = \epsilon^{ij}\frac{\partial \cdot}{\partial y^{i}} \frac{\partial 
     \cdot}{\partial y^{j}}. 
\end{equation}

We then have for $a\in W$
\begin{equation}
   {\cal{D}}[a,b]_{P} =   [{\cal{D}}a,b]_{P} +   [a,{\cal{D}}b]_{P}   
\end{equation}

We may construct abelian connections as before
\begin{equation}
    {\cal{D}}_{A} = D_{S} + [\frac{1}{h}\omega_{ij}y^{i}e^{j},\bullet]_{P} + [r,\bullet]_{P} 
\end{equation}
such that
\begin{equation}
 {\cal{D}}_{A}{\cal{D}}_{A}a = [\Omega ,a]_{P}=0 ,    
 \end{equation}
$\Omega$ being a central section of the Weyl bundle. We now deal with 
sections $a \in \Omega_{A}^{0}$ of the Weyl bundle obeying
\begin{equation}
 {\cal{D}}_{A} a = 0.
\end{equation}

Its explicit expression may be obtained as before. We have again a 
one-to-one correpondence between $a \in \Omega_{A}^{0}$ and its projection 
$a_{0}= \sigma a$. The projection of the Poisson bracket in $W$ now yields
\begin{equation}
\sigma [a,b]_{P} = \epsilon^{ij}\partial_{i}a_{0} \partial_{j}b_{0} 
\end{equation}
where $\sigma a= a_{0} $ and $\sigma b =b_{0}$.

We may now reconstruct Yang-Mills connections and extend them  in a 
global way over $\Sigma$. Everything follows exactly in the same way 
as before by changing the Weyl bracket by the Poisson one.
The corresponding supermembrane action may now be written 
in terms of the Poisson bracket. It is this canonical action that 
describes the doubly compactified $D=11$ Supermembrane as was first found in 
\cite{mor2}. It agrees with the noncommutative action obtained from the Weyl 
bracket when we expand up to degree 2. It must be in this way since the supermembrane action 
neither depends on the parameter $h$ of the formal deformation quantization nor 
on the arbitrary totally symmetric symbol present in the symplectic connection. 
In this sense the symplectic noncommutative action 
of the supermembrane in \cite{mor2}  when formulated in terms of the $star$ connection using a Seiberg-Witten map,
provides an action for a noncommutative star theory which does not depend 
either on $h$ or on the totally symmetric symbol in the symplectic connection.
It is natural to think that there is a one-to-one correspondence between the gauge equivalent 
classes of the Yang-Mills connections constructed from the Weyl 
bracket and with the Poisson one. We will discuss this relation and 
the corresponding Seiberg-Witten map in a forthcoming paper.

We notice that in both actions in addition to the standard 
noncommutative Yang-Mills terms, the integral of the projection of 
the curvature of the corresponding Yang-Mills connection on the Weyl 
bundle is present as well. This term characterizes the Weyl algebra bundle  as a vector bundle, 
it is introduced from the global construction which is expressed in terms of inner derivatives only , 
and there is no way to avoid it from this global point of view. This term can not be eliminated in the 
description of the supermembrane. Fundamental properties of the theory, 
such as the spectrum, change dramatically if it is fixed to zero.

\section{ Conclusions}
We have set up a global formulation for noncommutative Yang Mills theory 
over general symplectic manifolds using the concept of 
connections over Weyl algebra bundles on symplectic manifolds . This result is used to 
construct an action for a noncommutative supermembrane where a 
new topological term appears. We have found that the ordinary supermembrane 
action as known in the literature and the noncommutative supermembrane 
as constructed here may be obtained by special choices of 
connections in an Abelian Weyl bundle in each case. Also, that the 
framework of Weyl bundles suggests us to construct  actions 
for supermembranes and Yang Mills theories straightforwardly in a 
natural way whenever the symplectic structure is given from the 
start. We notice in the case of supermembranes that this symplectic 
structure may be provided by central charges in general. We took as 
example the case when the central charge is related to winding of 
coordinates in the target space in eleven dimensions for compact 
supermembranes. Similar results may be expected for the case of 
charged open membranes with boundaries.

\section* { {Acknowledgements}}
 
I. Martin and A. Restuccia thank the kind 
hospitality of the Imperial College's Theoretical Group and the King's 
College Mathematics Department, respectively, where this work was done.


\begin{thebibliography}{99}
    
    \bibitem{witnp1} E. Witten,  {\it Nucl. Phys.} {\bf B268} (1986) 253.
    \bibitem{doug} M. Douglas and N. Nekrasov, {\it Noncommutative 
    Field Theory} hep-th/0105 (2001)
    \bibitem{condou} A. Connes, M. Douglas and A. Schwarz, {\it JHEP}{\bf 02}:003 (1998). 
    \bibitem{witjh} N. Seiberg and E. Witten, {\it JHEP} {\bf 09}:032 (1999) 
    \bibitem{towns}  P.K. Townsend, {\it  Phys. Lett.} {\bf B350} (1995) 184;{\it  Phys. Lett.} {\bf B373} (1996) 68.
    \bibitem{konse} M. Kontsevich, ``Deformation quantization of Poisson manifolds'', q-alg/9709040 (1997)
    \bibitem{konscatt}A. Cattaneo and G. Felder,{\it  Commun. Math. 
    Phys. }{\bf 212} (2000) 591;  A. Cattaneo and G. Felder,{\it  
    Mod. Phys. Lett. }{\bf 212} (2001) 179.  
    \bibitem{mrt} I. Martin, A. Restuccia and R. Torrealba, {\it Nucl. Phys.} {\bf B521} (1998) 117.
    \bibitem{mor1}I. Martin, J. Ovalle and A. Restuccia, {\it  Phys. Lett.} {\bf B472} (2000) 77.  
    \bibitem{mor2}I. Martin, J. Ovalle and A. Restuccia, {\it  Phys. Rev.} {\bf D62} (2001)   
    \bibitem{mosq}F. Bayen, M. Flato, C. Fronsdal, A. Lichnerowicz 
    and D. Sternheimer, {\it Ann. Phys.} {\bf 111} (1978) 61.
    \bibitem{vey} J. Vey ,{\it  Comment. Math. Helv. }{\bf 50} (1975) 421.    
    \bibitem{fedo} B. Fedosov,{\it J. Diff. Geom.} {\bf 40} (1994) 213; 
    {\it  Deformation Quantization and Index Theory}, Akademie Verlag, 
    Berlin (1996).
    \bibitem{asak} T. Asakawa and I. Kishimoto,{\it Nucl. Phys.} {\bf 
    B591} (2000) 611.
    \bibitem{wess} B. Jurco, L. Moller, S. Schraml, P. Schupp and J. 
     Wess,{\it  Construction of non-abelian gauge theories on 
     noncommutative spaces}, hep-th/0104153 (2001).
    \bibitem{br2} Bellorin and A. Restuccia,  {\it  Phys. Rev.} {\bf D58} (2001)
    \bibitem{town2}  P.K. Townsend, {\it M theory from its 
    superalgebra}, hep-th/9712004 (1997).
 
\end{thebibliography}
\end{document}